\documentclass{llncs}
\usepackage[utf8]{inputenc}
\usepackage{url}
\usepackage[section]{placeins}
\usepackage{amssymb}
\usepackage{pi font}
\usepackage{microtype}
\usepackage{draftwatermark}
\SetWatermarkText{Pre-print}
\SetWatermarkScale{2}
\SetWatermarkText{\textsc{Pre-print}}

\begin{document}
\title{Integrated Safety and Security Risk Assessment Methods: A Survey of Key Characteristics and Applications}
\author{Sabarathinam Chockalingam\textsuperscript{1}, Dina Had\v{z}iosmanovi\'{c}\textsuperscript{2}, Wolter Pieters\textsuperscript{1}, Andr\'{e} Teixeira\textsuperscript{1}, and Pieter van Gelder\textsuperscript{1}}
\institute{\textsuperscript{1}Faculty of Technology, Policy and Management, Delft University of Technology, The
Netherlands\\\{S.Chockalingam, W.Pieters, Andre.Teixeira, P.H.A.J.M.vanGelder\}@tudelft.nl\\
\textsuperscript{2}Deloitte, The Netherlands\\
DHadziosmanovic@deloitte.nl}
\maketitle

\begin{abstract}
Over the last years, we have seen several security incidents that compromised system safety, of which some caused physical harm to people. Meanwhile, various risk assessment methods have been developed that integrate safety and security, and these could help to address the corresponding threats by implementing suitable risk treatment plans. However, an overarching overview of these methods, systematizing the characteristics of such methods, is missing. In this paper, we conduct a systematic literature review, and identify 7 integrated safety and security risk assessment methods. We analyze these methods based on 5 different criteria, and identify key characteristics and applications. A key outcome is the distinction between sequential and non-sequential integration of safety and security, related to the order in which safety and security risks are assessed. This study provides a basis for developing more effective integrated safety and security risk assessment methods in the future.\\ 


\end{abstract}

\section{Introduction}\label{sec:Introduction}
Information technologies and communication devices are increasingly being integrated into modern control systems \cite{Kriaa}. These modern control systems are used to operate life-critical systems where the human lives are at stake in case of failure. At the same time, they are often vulnerable to cyber-attacks, which may cause physical impact. An incident in Lodz is a typical example where a cyber-attack resulted in the derailment of 4 trams, and the injury of 12 people \cite{RISI}. It is therefore becoming increasingly important to address the combination of safety and security in modern control systems.
\par However, safety and security have been represented by separate communities in both academia and industry \cite{Stoneburner}. In our context, we think of the safety community as dealing with unintentional/non-malicious threats caused by natural disasters, technical failures, and human error. On the other hand, we think of the security community as dealing with intentional/malicious threats caused by intentional human behavior. 
\par Risk management plays a major role in dealing with both unintentional/non-malicious, and intentional/malicious threats. In the recent years, we have seen a transformation among the researchers of safety and security community to work together especially in risk management. As an example, there are developments of integrated safety and security risk assessment methods \cite{Macher,Schmittnerc,Sabaliauskaite,Schmittner,Chen,Steiner,Fovino}. Risk assessment is one of the most crucial parts of the risk management process as it is the basis for making risk treatment decisions \cite{ENISA}. The integrated safety and security risk assessment method helps to improve the completeness of risk assessment conducted by covering the interactions between malicious and non-malicious risks. However, a comprehensive review of integrated safety and security risk assessment methods which could help to identify their key characteristics and applications is lacking. Therefore, this research aims to fill this gap by addressing the research question: “What are the key characteristics of integrated safety and security risk assessment methods, and their applications?”. The research objectives are:
\begin{itemize}
\item[$\bullet$]\textbf{RO 1.} To identify integrated safety and security risk assessment methods. 
\item[$\bullet$]\textbf{RO 2.} To identify key characteristics and applications of integrated safety and security risk assessment methods based on the analysis of identified methods. 
\end{itemize}
\par The scope of this analysis covers important features of identified integrated safety and security risk assessment methods mainly, in terms of how these methods are created, and what the existing applications of these methods are. The analysis of identified methods is performed based on the following criteria: I. Citations in the Scientific Literature, II. Steps Involved, III. Stage(s) of Risk Assessment Process Addressed, IV. Integration Methodology, and V. Application(s) and Application Domain. The motivations for selecting these criteria are described in Section 5.  
\par The remainder of this paper is structured as follows: Section 2 describes the related work, followed by the review methodology in Section 3. In Section 4, we present the identified integrated safety and security risk assessment methods, and describe the steps involved in these methods. In Section 5, we perform the analysis of identified methods based on the criteria that we defined above. Finally, we highlight key characteristics and applications of integrated safety and security risk assessment methods followed by a discussion of future work directions in Section 6.

\section{Related Work}\label{sec:Related Work}
Cherdantseva et al. presented 24 cybersecurity risk assessment methods for Supervisory Control and Data Acquisition (SCADA) systems \cite{Cherdantseva}. In addition, they analyzed the presented methods based on the following criteria: I. Aim, II. Application domain, III. Stages of risk management addressed, IV. Key concepts of risk management covered, V. Impact measurement, VI. Sources of data for deriving probabilities, VII. Evaluation method, and VIII. Tool support. Based on the analysis, they suggested the following categorization schemes: I. Level of detail and coverage, II. Formula-based vs. Model-based, III. Qualitative vs. Quantitative, and IV. Source of probabilistic data. However, Cherdantseva et al. did not present integrated safety and security risk assessment methods. We used and complemented some of the criteria provided by Cherdantseva et al. to perform the analysis of integrated safety and security risk assessment methods as described in Section 5. 
\par Risk assessment methods like Failure Mode and Effects Analysis (FMEA) \cite{IEC}, Fault Tree Analysis (FTA) \cite{Lee}, Component Fault Tree (CFT) \cite{Kaiser} have been used by safety community whereas the risk assessment methods like Attack Trees \cite{Schneier}, Attack-Countermeasure Trees (ACT) \cite{Roy}, National Institute of Standards and Technology (NIST) 800-30 Risk Assessment \cite{NIST} have been used by security community. Several authors used these methods as a starting point for the development of integrated safety and security risk assessment methods.
\par Kriaa et al. highlighted standard initiatives such as ISA-99 (Working Group 7), IEC TC65 (Ad Hoc Group 1), IEC 62859, DO-326/ED-202 that consider safety and security co-ordination for Industrial Control Systems (ICS) \cite{Kriaa}. They described various generic approaches that considered safety and security at a macroscopic level of system design or risk evaluation, and also model-based approaches that rely on a formal or semi-formal representation of the functional/non-functional aspects of system. They classified the identified approaches based on the following criteria: I. Unification vs. Integration, II. Development vs. Operational, and III. Qualitative vs. Quantitative. However, Kriaa et al. did not primarily focus on integrated safety and security risk assessment methods that have been already applied in at least one real-case/example involving control system. Also, Kriaa et al. did not identify key characteristics and applications of integrated safety and security risk assessment methods. We included methods such as Failure Mode, Vulnerabilities, and Effect Analysis (FMVEA) \cite{Schmittner}, Extended Component Fault Tree (CFT) \cite{Steiner}, and Extended Fault Tree (EFT) \cite{Fovino} from Kriaa et al. in our work as they satisfy our selection criteria. In addition, we included other methods that satisfy our selection criteria, such as Security-Aware Hazard Analysis and Risk Assessment (SAHARA) \cite{Macher}, Combined Harm Assessment of Safety and Security for Information Systems (CHASSIS) \cite{Schmittnerc}, Failure-Attack-CountTermeasure (FACT) Graph \cite{Sabaliauskaite}, and Unified Security and Safety Risk Assessment \cite{Chen}.

\section{Review Methodology}\label{sec:Review Methodology}
This section describes the methodology for selecting the integrated safety and security risk assessment methods. The selection of these methods mainly consists of two stages:
\begin{itemize}
\item[$\bullet$]Searches were performed on IEEE Xplore Digital Library, ACM Digital Library, Scopus, DBLP, and Web of Science – All Databases. The search-strings were constructed from keywords “Attack”, “Failure”, “Hazard”, “Integration”, “Risk”, “Safety”, “Security”, and “Threat”. DBLP provided a good coverage of relevant journals and conferences.
\item[$\bullet$]Methods were selected from the search results according to the following criteria: 
\subitem The method should address any or all of the following risk assessment stages: risk identification, risk analysis, and/or risk evaluation.
\subitem The method should consider both unintentional and intentional threats.
\subitem The method should have been already applied in at least one real-case/\\example involving control system.
\subitem The literature should be in English language.
\end{itemize}
\par Once an integrated safety and security risk assessment method was selected, the scientific literature that cited it was also traced.

\section{Integrated Safety and Security Risk Assessment Methods}\label{sec:Integrated Safety and Security Risk Assessment Methods}
This section presents the identified integrated safety and security risk assessment methods, and describes the steps involved in these methods. This section aims to address the RO 1. Based on the review methodology described in Section 3, we have identified 7 integrated safety and security risk assessment methods: I. SAHARA \cite{Macher}, II. CHASSIS \cite{Schmittnerc}, III. FACT Graph \cite{Sabaliauskaite}, IV. FMVEA \cite{Schmittner}, V. Unified Security and Safety Risk Assessment \cite{Chen}, VI. Extended CFT \cite{Steiner}, and VII. EFT \cite{Fovino}.  
\subsection{SAHARA Method}
The steps involved in the SAHARA method \cite{Macher} are as follows: I. The ISO 26262 – Hazard Analysis and Risk Assessment (HARA) approach is used in a conventional manner to classify the safety hazards according to the Automotive Safety Integrity Level (ASIL), and to identify the safety goal and safe state for each identified potential hazard; II. The attack vectors of the system are modelled. The STRIDE method is used to model the attack vectors of the system \cite{Macher,Scandariato}; III. The security threats are quantified according to the Required Resources (R), Required Know-how (K), and Threat Criticality (T); IV. The security threats are classified according to the Security Level (SecL). SecL is determined based on the level of R, K, and T; V. Finally, the security threats that may violate the safety goals (T$>$2) are considered for the further safety analysis.
\subsection{CHASSIS Method}
The steps involved in the CHASSIS method \cite{Schmittnerc} are as follows: I. The elicitation of functional requirements which involve creating the use-case diagrams that incorporates the users, system functions and services; II. The elicitation of safety and security requirements which involve creating misuse case diagram based on the identified scenarios for safety and security involving faulty-systems and attackers respectively; III. Trade-off discussions are used to support the resolution of conflict between the safety, and security mitigations.
\subsection{FACT Graph Method}
The steps involved in the FACT Graph method \cite{Sabaliauskaite} are as follows: I. The fault trees of the system analyzed are imported to start the construction of FACT graph; II. The safety countermeasures are attached to the failure nodes in the FACT graph; III. The attack trees of the system analyzed are imported to the FACT graph in construction. This is done by adding an attack-tree to the failure node in the FACT graph with the help of OR gate, if the particular failure may also be caused by an attack; IV. The security countermeasures are attached to the attack nodes in the FACT graph. This could be done based on the ACT technique \cite{Roy}.
\subsection{FMVEA Method}
The steps involved in the FMVEA method \cite{Schmittner} are as follows: I. A functional analysis at the system level is performed to get the list of system components; II. A component that needs to be analyzed from the list of system components is selected; III. The failure/threat modes for the selected component are identified; IV. The failure/threat effect for each identified failure/threat mode is identified; V. The severity for the identified failure/threat effect is determined; VI. The potential failure causes/vulnerabilities/threat agents are identified; VII. The failure/attack probability is determined. Schmittner et al. described the attack probability as the sum of threat properties and system susceptibility ratings. The threat properties is the sum of motivation and capabilities ratings, whereas the system susceptibility is the sum of reachability and unusualness of the system ratings; VIII. Finally, the risk number is determined, which is the product of severity rating and failure/attack probability.
\subsection{Unified Security and Safety Risk Assessment Method}
The steps involved in the Unified Security and Safety Risk Assessment method \cite{Chen} are as follows: I. The system boundary, system functions, system and data criticality, system and data sensitivity are identified; II. The threats, hazards, vulnerabilities, and hazard-initiating events are identified; III. The current and planned controls are identified; IV. The threat likelihood is determined; V. The hazard likelihood is determined; VI. The asset impact value is determined; VII. The combined safety-security risk level is determined; VIII. The control recommendations are provided; IX. The risk assessment reports are provided.
\subsection{Extended CFT Method}
The steps involved in the extended CFT method \cite{Steiner} are as follows: I. The CFT for the system analyzed is developed. This could be done based on \cite{Kaiser}; II. The CFT is extended by adding an attack tree to the failure node with the help of OR gate, if the particular event may also be caused by an attack; III. The qualitative analysis is conducted by calculating Minimal Cut Sets (MCSs) per top level event. MCSs containing only one event would be single point of failure which should be avoided; IV. The quantitative analysis is conducted by assigning values to the basic events. Therefore, MCSs containing only safety events would have a probability P, MCSs containing only security events would have a rating R, MCSs containing both safety and security events would have a tuple of probability and rating (P, R).
\subsection{EFT Method}
The steps involved in the EFT method \cite{Fovino} are as follows: I. The fault tree for the system analyzed is developed by taking into account the random faults; II. The developed fault tree is extended by adding an attack tree to the basic or intermediate event in the fault tree, if the particular event in the fault tree may also be caused by malicious actions. The attack tree concept used in the development of EFT is based on \cite{IFovino}; III. The quantitative analysis is performed based on the formulae defined in \cite{Fovino} which help to calculate the top event probability.

\section{Analysis of Integrated Safety and Security Risk Assessment Methods}\label{sec:Analysis of Integrated Safety and Security Risk Assessment Methods}
This section performs the analysis of integrated safety and security risk assessment methods based on the criteria: I. Citations in the Scientific Literature, II. Steps Involved, III. Stage(s) of Risk Assessment Process Addressed, IV. Integration Methodology, and V. Application(s) and Application Domain. This allows us to identify key characteristics and applications of integrated safety and security risk assessment methods. This section aims to address the RO 2.
\par The integrated safety and security risk assessment methods described in the previous section are listed in Table 1. In Table 1, country is the country of the first author of the paper and citations is the number of citations of the paper according to Google Scholar Citation Index as on 31st August 2016.
\begin{table}[ht]
\centering
\caption{List of Integrated Safety and Security Risk Assessment Methods (Ordered by the number of citations)}
\begin{tabular}{|p{7.8cm}|p{1cm}|p{1.5cm}|p{1.5cm}|}
\hline
\textbf{Integrated Safety and Security Risk Assessment Method}&\textbf{Year}&\textbf{Country}&\textbf{Citations}\\
\hline
EFT \cite{Fovino}&2009&Italy&63\\
\hline
Extended CFT \cite{Steiner}&2013&Germany&17\\
\hline 
FACT Graph \cite{Sabaliauskaite}&2015&Singapore&5\\
\hline 
CHASSIS \cite{Schmittnerc}&2015&Austria&4\\
\hline
FMVEA \cite{Schmittner}&2014&Austria&4\\
\hline
SAHARA \cite{Macher}&2015&Austria&2\\
\hline
Unified Security and Safety Risk Assessment \cite{Chen}&2014&Taiwan&1\\
\hline
\end{tabular}
\end{table}
\par From Table 1, we observe that the researchers started to recognize the importance of integrated safety and security risk assessment methods which resulted in the increase in number of papers produced especially during 2014, and 2015. The largest number of citations (63) is acquired by the EFT method published in 2009. The second most cited paper, among analyzed, with 17 citations, is the Extended CFT method published in 2013. However, it is understandable that the methods published during the last few years received lower number of citations ranging from 1 to 5.
\par Based on the steps involved in each method as described in Section 4, we conclude that there are two types of integrated safety and security risk assessment methods:
\begin{itemize}
\item[$\bullet$]Sequential Integrated Safety and Security Risk Assessment Method: In this type of method, the safety risk assessment, and security risk assessment are performed in a particular sequence. For instance, the Extended CFT method starts with the development of CFT for the system analyzed. Later, the attack tree is added to extend the developed CFT. This method starts with the safety risk assessment followed by the security risk assessment. Methods such as SAHARA, FACT Graph, Unified Security and Safety Risk Assessment, Extended CFT, and EFT come under the sequential type.
\item[$\bullet$]Non-sequential Integrated Safety and Security Risk Assessment Method: In this type of method, the safety risk assessment, and security risk assessment are performed without any particular sequence. For instance, in the FMVEA method, the results of safety risk assessment and security risk assessment are tabulated in the same table without any particular sequence. Methods such as FMVEA and CHASSIS come under the non-sequential type.
\end{itemize}
\par Cherdantseva et al. used ‘stage(s) of risk management process addressed’ as a criteria to analyze the identified cybersecurity risk assessment methods for SCADA systems \cite{Cherdantseva}. We adapted and used this criteria as ‘stage(s) of risk assessment process addressed’ because the major focus of our research is on risk assessment. This criteria will allow us to identify the predominant stage(s) of risk assessment process addressed by the integrated safety and security risk assessment methods.
\par A risk assessment process consists of typically three stages: 
\begin{itemize}
\item [$\bullet$]Risk Identification: This is the process of finding, recognizing and describing the risks \cite{ISO}. 
\item [$\bullet$]Risk Analysis: This is the process of understanding the nature, sources, and causes of the risks that have been identified and to estimate the level of risk \cite{ISO}.
\item [$\bullet$]Risk Evaluation: This is the process of comparing risk analysis results with risk criteria to make risk treatment decisions \cite{ISO}.
\end{itemize}
\par Table 2 highlights the integrated safety and security risk assessment method and the corresponding stage(s) of the risk assessment process addressed. This is done based on the definitions of risk identification, risk analysis, and risk evaluation. We also take into account the safety risk assessment method, and security risk assessment method that were combined in the integrated safety and security risk assessment method.\\
\begin{table}[ht]
\centering
\caption{Stage(s) of Risk Assessment Process Addressed}
\begin{tabular}{|p{6.4cm}|p{2.2cm}|p{1.4cm}|p{1.8cm}|}
\hline
\textbf{Integrated Safety and Security Risk\newline Assessment Method}&\textbf{Risk\newline Identification}&\textbf{Risk\newline Analysis}&\textbf{Risk\newline Evaluation}\\
\hline
SAHARA&\checkmark&\checkmark&$\times$\\
\hline
CHASSIS&\checkmark&$\times$&$\times$\\
\hline 
FACT Graph&\checkmark&$\times$&$\times$\\
\hline
FMVEA&\checkmark&\checkmark&$\times$\\
\hline 
Unified Security and Safety Risk Assessment&\checkmark&\checkmark&\checkmark\\
\hline
Extended CFT&\checkmark&\checkmark&$\times$\\
\hline
EFT&\checkmark&\checkmark&$\times$\\ 
\hline
\end{tabular}
\raggedright{In Table 2, {\checkmark}($\times$) indicates that the particular method addressed (did not address) the corresponding risk assessment stage.}
\end{table}
\par From Table 2, we understand that all methods addressed the risk identification, 5 out of 7 methods addressed the risk analysis, whereas only 1 out of 7 methods addressed the risk evaluation stage of the risk assessment process. This implies that the risk evaluation stage is not given much attention compared to the other stages of the risk assessment process in the integrated safety and security risk assessment methods. Cherdantseva et al. also highlighted that the majority of the cybersecurity risk assessment methods for SCADA systems concentrates on the risk identification and risk analysis stages of the risk assessment process \cite{Cherdantseva}.
\par We used the criteria ‘Integration methodology’ because this will allow us to understand which combination of safety, and security risk assessment methods are being used in the integrated safety and security risk assessment methods as summarized in Table 3.
\begin{table}[ht]
\centering
\caption{Integration Methodology}
\begin{tabular}{|p{3.8cm}|p{4cm}|p{4cm}|}
\hline
\textbf{Integrated Safety and Security Risk\newline Assessment Method}&\textbf{Safety Risk Assessment Method}&\textbf{Security Risk\newline Assessment Method}\\
\hline
SAHARA & ISO 26262: HARA & Variation of ISO 26262:\newline HARA \\ 
\hline
CHASSIS & Safety Misuse Case (Involving Faulty-systems) & Security Misuse Case (Involving Attackers) \\
\hline 
FACT Graph & Fault Tree & Attack Tree \\ 
\hline
FMVEA & FMEA & Variation of FMEA \\ 
\hline 
Unified Security and Safety Risk Assessment & Variation of NIST 800-30 Security Risk Estimation & NIST 800-30 Security Risk Estimation \\
\hline
Extended CFT & CFT & Attack Tree \\
\hline
EFT & Fault Tree & Attack Tree \\
\hline 
\end{tabular}
\end{table}

\par From Table 3, we observe that there are four ways in which the integrated safety and security risk assessment methods have been developed:
\begin{itemize}
\item[$\bullet$]Integration through the combination of a conventional safety risk assessment method and a variation of the conventional safety risk assessment method for security risk assessment. The methods SAHARA and FMVEA come under this category.
\item[$\bullet$]Integration through the combination of a conventional security risk assessment method and a variation of the conventional security risk assessment method for safety risk assessment. The Unified Security and Safety Risk Assessment method come under this category.
\item[$\bullet$]Integration through the combination of a conventional safety risk assessment method and a conventional security risk assessment method. The methods FACT Graph, Extended CFT, and EFT come under this category.
\item[$\bullet$]Others - There is no conventional safety risk assessment, and conventional security risk assessment method used in the integration. The CHASSIS method come under this category. The CHASSIS method used a variation of Unified Modeling Language (UML)-based models for both the safety and security risk assessment.
\end{itemize} 
\par We used the criteria ‘Application(s) and Application domain’ because this will allow us to understand the type of application(s), and the corresponding application domain of integrated safety and security risk assessment methods. Table 4 highlights the integrated safety and security risk assessment method and the corresponding application(s) and application domain. 

\begin{table}[ht]
\centering
\caption{Application(s) and Application Domain}
\begin{tabular}{|p{3cm}|p{6cm}|p{2.8cm}|}
\hline
\textbf{Integrated Safety and Security Risk Assessment Method}&\textbf{Application(s)}&\textbf{Application\newline Domain}\\
\hline
SAHARA&Battery Management System use-case \cite{Macher}&Transportation\\
\hline
CHASSIS&Over The Air (OTA) system \cite{Schmittnerc}, Air traffic management remote tower example \cite{Raspotnig}.&Transportation\\
\hline
FACT Graph&Over-pressurization of a vessel example \cite{Sabaliauskaite}&Power and Utilities\\ 
\hline
FMVEA&OTA system [5], Telematics control unit \cite{Schmittner}, Engine test-stand \cite{Schmittnerch}, Communications-based train control system \cite{Chenb}.&Transportation\\
\hline
Unified Security and Safety Risk\newline Assessment&High pressure core flooder case-study \cite{Chen}&Power and Utilities\\
\hline
Extended CFT&Adaptive cruise control system \cite{Steiner}&Transportation\\
\hline
EFT&Release of toxic substance into the environment example \cite{Fovino}&Chemical\\
\hline
\end{tabular}
\end{table}

\par From Table 4, we observe that 4 methods were applied in the transportation domain, 2 methods were applied in the power and utilities domain, and 1 method was applied in the chemical domain. The major development, and application of integrated safety and security risk assessment methods, is in the transportation domain. The Threat Horizon 2017 listed “death from disruption to digital services” as one of the threats especially in the transportation and medical domain \cite{ISF}. In the transportation domain, there is a potential for cyber-attacks which compromises system safety and result in the injury/death of people which was illustrated by a tram incident in Lodz \cite{RISI}.

\section{Conclusions and Future Work}\label{sec:Conclusions and Future Work}
In this paper, we have identified 7 integrated safety and security risk assessment methods. Although we cannot completely rule out the existence of other unobserved integrated safety and security risk assessment methods that fulfil our selection criteria, the review methodology that we adopted helped to ensure the acceptable level of completeness in the selection of these methods. Based on the analysis, we identified key characteristics and applications of integrated safety and security risk assessment methods. 
\begin{itemize}
\item[$\bullet$]There are two types of integrated safety and security risk assessment methods based on the steps involved in each method. They are: a. Sequential, and b. Non-sequential.    
\item[$\bullet$]There are four ways in which the integrated safety and security risk assessment methods have been developed. They are: a. The conventional safety risk assessment method as the base and a variation of the safety risk assessment method for security risk assessment, b. The conventional security risk assessment method as the base and a variation of the security risk assessment method for safety risk assessment, c. A combination of a conventional safety risk assessment method, and a conventional security risk assessment method, d. Others.
\item[$\bullet$]Risk identification and risk analysis stages were given much attention compared to the risk evaluation stage of the risk assessment process in the integrated safety and security risk assessment methods. 
\item[$\bullet$]Transportation, power and utilities, and chemical were the three domains of application for integrated safety and security risk assessment methods.
\end{itemize}
\par The identified integrated safety and security risk assessment methods did not take into account real-time system information to perform dynamic risk assessment which needs to be addressed to make it more effective in the future. This study provided the list of combinations of safety, and security risk assessment methods used in the identified integrated safety and security risk assessment methods. In the future, this would act as a base to investigate the other combinations of safety, and security risk assessment methods that could be used in the development of more effective integrated safety and security risk assessment methods. Furthermore, this study provided the type of applications and application domains of the identified integrated safety and security risk assessment methods. In the future, this would act as a starting point to evaluate the applicability of these methods in the other domains besides transportation, power and utilities, and chemical.

\section*{Acknowledgements}\label{Acknowledgements}
This research received funding from the Netherlands Organisation for Scientific Research (NWO) in the framework of the Cyber Security research program. This research has also received funding from the European Union's Seventh Framework Programme (FP7/2007-2013) under grant agreement ICT-318003 (TREsPASS). This publication reflects only the authors' views and the Union is not liable for any use that may be made of the information contained herein.


\begin{thebibliography}{1}

\bibitem{Kriaa}
Kriaa, S., Pietre-Cambacedes, L., Bouissou, M., Halgand, Y.: A Survey of Approaches Combining Safety and Security for Industrial Control Systems. Reliability Engineering \& System Safety. vol. 139, pp. 156 – 178. (2015)

\bibitem{RISI}
RISI Database.: Schoolboy Hacks into Polish Tram System (2016).\url{http://www.risidata.com/Database/Detail/schoolboy_hacks_into_polish_tram_system}

\bibitem{Stoneburner}
Stoneburner, G.: Toward a Unified Security-Safety Model. Computer. vol. 39, no. 8, pp. 96 – 97. (2006)

\bibitem{Macher} 
Macher, G., H\"{o}ller, A., Sporer, H., Armengaud, E., Kreiner, C.: A Combined Safety – Hazards and Security - Threat Analysis Method for Automotive Systems. Koornneef, F., van Gulijk, C. (eds.) SAFECOMP 2015 Workshops. LNCS, vol. 9338, pp. 237 – 250. Springer, Heidelberg (2015)

\bibitem{Schmittnerc}
Schmittner, C., Ma, Z., Schoitsch, E., Gruber, T.: A Case Study of FMVEA and CHASSIS as Safety and Security Co-Analysis Method for Automotive Cyber Physical Systems. In: Proceedings of the 1st ACM Workshop on Cyber Physical System Security (CPSS), pp. 69 – 80. (2015)

\bibitem{Sabaliauskaite}
Sabaliauskaite, G., Mathur, A.P.: Aligning Cyber-physical System Safety and Security. Cardin, M.A., Krob, D., Cheun, L.P., Tan, Y.H., Wood, K. (eds.) Complex Systems Design \& Management Asia 2014. LNCS, pp. 41 – 53. (2015)

\bibitem{Schmittner}
Schmittner, C., Ma, Z., Smith, P.: FMVEA for Safety and Security Analysis of Intelligent and Cooperative Vehicles. Bondavalli, A., Ceccarelli, A., Ortmeier, F. (eds.) SAFECOMP 2014 Workshops. LNCS, vol. 8696, pp. 282 – 288. Springer, Heidelberg (2014)

\bibitem{Chen}
Chen, Y., Chen, S., Hsiung, P., Chou, I.: Unified Security and Safety Risk Assessment – A Case Study on Nuclear Power Plant. In: Proceedings of the International Conference on Trusted Systems and their Applications (TSA), pp. 22 – 28. (2014)

\bibitem{Steiner}
Steiner, M., Liggesmeyer, P., Combination of Safety and Security Analysis – Finding Security Problems that Threaten the Safety of a System. In: Workshop on Dependable Embedded and Cyber-physical Systems (DECS), pp. 1 – 8. (2013)

\bibitem{Fovino}
Fovino, I.N., Masera, M., De Cian, A., Integrating Cyber Attacks within Fault Trees. Reliability Engineering and System Safety. vol. 94, no. 9, pp. 1394 – 1402. (2009)

\bibitem{ENISA}
European Union Agency for Network and Information Security (ENISA). The Risk Management Process (2016). \url{https://www.enisa.europa.eu/activities/risk-management/current-risk/risk-management-inventory/rm-process}

\bibitem{Cherdantseva}
Cherdantseva, Y., Burnap, P., Blyth, A., Eden, P., Jones, K., Soulsby, H., Stoddart, K.: A Review of Cyber Security Risk Assessment Methods for SCADA Systems. Computers \& Security. vol. 56, pp. 1 – 27. (2016)

\bibitem{IEC}
International Electrotechnical Commission (IEC).: IEC 60812: Analysis Techniques for System Reliability – Procedures for Failure Mode and Effects Analysis. (2006)

\bibitem{Lee}
Lee, W.S., Grosh, D.L., Tillman, F.A., Lie, C.H.: Fault Tree Analysis, Methods, and Applications – A Review. IEEE Transactions on Reliability. vol. R-34, no. 3, pp. 194 – 203. (1985)

\bibitem{Kaiser}
Kaiser, B., Liggesmeyer, P., Mackel, O.: A New Component Concept for Fault Trees. In: Proceedings of the 8th Australian Workshop on Safety Critical Systems and Software (SCS), vol. 33, pp. 37 – 46. (2003)

\bibitem{Schneier}
Schneier, B.: Attack Trees. Dr. Dobb’s Journal. vol. 24, no. 12, pp. 21 – 29. (1999)

\bibitem{Roy}
Roy, A., Kim, D.S., Trivedi, K.S.: Scalable Optimal Countermeasure Selection Using Implicit Enumeration on Attack Countermeasure Trees. In: Proceedings of the 42nd Annual IEEE/IFIP International Conference on Dependable Systems and Networks (DSN), pp. 1 – 12. (2012)

\bibitem{NIST}
National Institute of Standards and Technology (NIST).: Risk Management Guide for Information Technology Systems. (2002)

\bibitem{Scandariato}
Scandariato, R., Wuyts, K., Joosen, W.: A Descriptive Study of Microsoft’s Threat Modeling Technique. Requirements Engineering. vol. 20, no. 2, pp. 163-180. (2015)

\bibitem{IFovino}
Fovino, I.N., Masera, M.: Through the Description of Attacks: A Multi-Dimensional View. Gorski, J. (eds.) SAFECOMP 2006. LNCS, vol. 4166, pp. 15 – 28. Springer, Heidelberg (2006)

\bibitem{ISO}
International Organisation for Standardization (ISO).: ISO 31000: 2009 - Risk Management – Principles and Guidelines. (2009)

\bibitem{ISF}
Information Security Forum.: Threat Horizon 2017: Dangers Accelerate (2015). \url{https://www.securityforum.org/uploads/2015/03/Threat-Horizon_2017_Executive-Summary.pdf}

\bibitem{Raspotnig}
Raspotnig, C., Karpati, P., Katta, V.: A Combined Process for Elicitation and Analysis of Safety and Security Requirements. Bider, I., Halpin, T., Krogstie, J., Nurcan, S., Proper, E., Schmidt, R., Soffer, P., Wrycza, S. (eds.) BPMDS and EMMSAD 2012. LNBIP, vol. 113, pp. 347 – 361. Springer, Heidelberg (2012)

\bibitem{Schmittnerch}
Schmittner, C., Gruber, T., Puschner, P., Schoitsch, E.: Security Application of Failure Mode and Effect Analysis (FMEA). Bondavalli, A., Di Giandomenico, F. (eds.) SAFECOMP 2014. LNCS, vol. 8666, pp. 310 – 325. (2014) 

\bibitem{Chenb}
Chen, B., Schmittner, C., Ma, Z., Temple, W.G., Dong, X., Jones, D.L., Sanders, W.H.: Security Analysis of Urban Railway Systems: The Need for a Cyber-Physical Perspective. Koornneef, F., van Gulijk, C. (eds.) SAFECOMP 2015 Workshops. LNCS, vol. 9338, pp. 277 – 290. Springer, Heidelberg (2015)
   
\end{thebibliography}
\end{document}